\documentclass[preprint,amsmath,titlepage,amssymb,english,aps,showkeys]{revtex4}
\usepackage{xcolor}
\usepackage{amssymb,amsmath}
\usepackage{graphicx}
\usepackage{hyperref}
\usepackage[utf8]{inputenc}
\usepackage[T1]{fontenc}
\usepackage{amsmath}
\usepackage{amsfonts}

\begin{document}

\title{On the Stability of Non-Singular Solutions in Effective Theory from Kaluza-Klein Unimodular Gravity}

\author{J{\'u}lio C. Fabris}
\email{julio.fabris@cosmo-ufes.org}
\affiliation{Núcleo Cosmo-ufes\&Departamento de F{\'i}sica, CCE, Universidade Federal do Espírito Santo, Vitória, ES, Brazil}

\author{ Richard Kerner}
\email{richard.kerner@sorbonne-universite.fr}
\affiliation{Laboratoire de Physique Th\'eorique de la Mati\`ere Condens\'ee, Sorbonne-Universit\'e, Boîte 121, 4 Place Jussieu, 75005, Paris,  France}

\begin{abstract}

Unimodular theory incorporating the Kaluza-Klein construction in five dimensions leads, after reduction to four dimensions, to 
a new class of scalar-tensor theory. The vacuum cosmological solutions display a bouncing, non singular behavior. From the four dimensional point of view,
the solutions are completely regular. However, the propagation of gravitational waves in this geometry displays the presence of instabilities which reflect
some features of the original five dimensional structure. Comparison with a four dimensional quantum model with cosmological constant, which has a similar 
background behavior, is discussed.
\end{abstract}


\maketitle

\section{Introduction}

The Unimodular Gravity, was considered already in $1919$ by Einstein \cite{Einstein1919}, who investigated an alternative to the initial 
formulation of General Relativity by imposing the extra unimodularity condition on the pseudo-Riemannian metric, $\mid {\rm det} (g_{\mu \nu} \mid = 1$.
This condition seemed natural for at leat two reasons: firstly, the Minkowskian metric tensor $\eta_{\mu \nu} = {\rm diag} (+1,-1,-1,-1)$
has this property, and its infinitesimal deformations $h_{\mu \nu}$ should be traceless if the unimodularity was to be also imposed on 
the deformed metric $g_{\mu \nu} + \epsilon h_{\mu \nu}$; secondly, it permitted to replace the cosmological term by the Lagrange multiplier 
taking into account this constraint in the variational principle. In $1919$ the Friedmann solution was yet unknown, and Einstein was 
attached to the Aristotelean vision of an eternal and stationary (at large) Universe. This is why he introduced the cosmological term $\Lambda g_{\mu \nu}$
on the right-hand side of the field equations, acting as a source of negative pressure ensuring the balance with non-zero matter density and
its gravity. 

In his quest for a Unified Field Theory Einstein took interest in Th. Kaluza's $5$-dimensional generalizaton of General Relativity \cite{Kaluza}, especially after O. Klein's
article \cite{Klein} combining Kaluza's model with quantum mechanical interpretation of discrete electric charge provided by the $S^1$ topology of the 
fifth dimension. Einstein published his first comment on the five-dimensional Kaluza-Klein model as early as in $1927$ \cite{Einstein1927}, and a more substantial investigation
co-authored with P. Bergmann \cite{Einstein1938}, looking for possible particle-like solutions, which was not conclusive. Solutions of the monopole type were found on the Kaluza-Klein
manifolds much later, by Sorkin \cite{Sorkin} and Gross and Perry \cite{GrossPerry}. 

It is worthwhile to recall the so-called "Kaluza-Klein miracle", which is a happy coincidence leading to the possibility of neglecting the scalar field
component and considering solutions with only pure gravity and electromagnetic fields. In fact, a $5$-dimensional metric tensor $g_{AB}, \; \; A, B = 1,2,...5$
contains $15$ independent functions, and the generalized Einstein's system should consist in $15$ independent equations corresponding to symmetric
combinations of indices $AB$ in ${\tilde{G}}_{AB} = {\tilde{R}}_{AB} - \frac{1}{2} {\tilde{g}}_{AB} {\tilde{R}}$ (the tilde means that the corresponding
geometrical objects are evaluated in $5$ dimensions). Then we have $15$ combinations, $(\mu \nu), \; \; (\mu 5)$ and $(55)$, the last one yielding the 
equation of motion of the scalar field. By arbitrarily suppressing the scalar field component we are faced with $15$ independent equations for only
$14$ unknown functions, the $4D$ metric tensor $g_{\mu \nu}$ and the electromagnetic potential $A_{\mu}$; however, even with this forced ansats, 
the last equation $G_{55} = 0$ reduces to the tautology $0=0$. Much later, the scalar-tensor theory proposed by Brans and Dicke \cite{Brans} turned out to be equivalent
with the full Kaluza-Klein model including the scalar component $g_{55}$.
 
Another quite amazing thing is that during more than half a century since the advent of the Kaluza-Klein $5$-dimensional generalization of Einstein's 
General Relativity, the fact that in more than four dimensions other invariants of Riemannian curvature can be added to the variational principle from which the equations are
derived. In all generalizations and modifications of the Kaluza-Klein model, including the non-abelian generalizations with an arbitrary gauge Lie group
replacing the $1$-dimensional $U(1)$ symmetry of the electromagnetic interaction, the same Einstein-Hilbert variational principle was considered, the
integrand being the Riemann scalar density $R \; \sqrt{\mid g \mid}$. And this in spite of the fact that the invariants of higher degree were known, 
(see \cite{Lanczos}, \cite{Lovelock}), the next after scalar curvature being the Gauss-Bonnet invariant given by the formula 
$I_2 = (R_{ijkl}R^{ijkl} - 4\; R_{ik}R^{ik} + R^2 ) \; \sqrt{\mid g \mid}$. Even the seminal paper on Kaluza-Klein theory authored by E. Witten in $1981$
(\cite{Witten1981}) does not mention this possibility, which was taken into account in the context of $5$-dimensional version leading to a simplified
formulation of non-linear electrodynamics in \cite{Kerner1987}, and in cosmology using a $10$-dimensional non-abelian version in $1988$ \cite{Giorgini1988}

Another development in generalizations of Kaluza-Klein theories that was until recently overlooked - at least to our knowledge - is the application of unimodularity
to the multi-dimensional Kaluza-Klein metric tensor, and the consequences such a condition may bring to the gravitational sector after dimensional reduction.
The combination of the unimodularity constraint and the non-abelian version of the Kaluza-Klein model can bring new vistas not only to the multidimensional cosmological models,
based on the non-abelian generalizations, but also to the gauge field content and their interaction with gravity.
 
A formulation of Unimodular Gravity in five dimensions, using the Kaluza-Klein construction, has been carried out in Ref. \cite{Kerner}, 
and its implications for cosmological models and gravitational waves are the main subject of the present article.

\section{Preliminaries}

One of the most important problems in contemporary theoretical physics is the necessity of inclusion of exotic fluids that constitute the dark sector 
of the matter content of the universe. 
The two components of the dark sector, dark matter and dark energy, resist up to now to all tentatives of direct detection, while many indirect 
evidences suggest that they must exist, being responsible for the almost $95\%$ of the cosmic content. The dark energy in particular is associated 
with cosmological constant taking supposedly its origin in the vacuum energy density that results from quantum field-theoretical considerations \cite{jm}. 
However, the estimated value obtained from the current version of quantum field theory is by many dozens of orders 
of magnitude larger than its observed value, perhaps the worst discrepancy ever encountered in physics.

One possibility of interpretation of this discrepancy is that a degravitating mechanism may drastically reduce the observed gravitational effects 
of the vacuum energy. If such a mechanism exists, it could solve many puzzles at once. For example, if dark energy is described by a 
self-interacting quintessence scalar field, the gravitational contribution of the vacuum energy may be reduced to zero, and the cosmological constant may disappear 
from the gravitational equations. 

The degravitation of cosmological constant can be represented by a simple classical mechanism in Unimodular Gravity (UG). Originally, UG was proposed a few years after the formulation 
of the General Relativity (GR) theory by imposing a coordinate system such that $\sqrt{-g} = 1$. The consequence of such restriction is a set of traceless 
of gravitational field equations, which read
\begin{eqnarray}
\label{uge}
R_{\mu\nu} - \frac{1}{4}g_{\mu\nu}R = 8\pi G\biggr\{T_{\mu\nu} - \frac{1}{4}g_{\mu\nu}T\biggl\},
\end{eqnarray}
where $T_{\mu\nu}$ is the energy-momentum tensor, $T$ being its trace. One verifies directly that any contribution corresponding to cosmological constant 
automatically disappears. 

Nevertheless, the structure of equations (\ref{uge}) contains a "hidden" cosmological term. In fact, the imposition of the unimodular constraint 
breaks the general diffeomorphism invariance, a cornestone of GR theory, leading to the transverse diffeomorphism. Hence, the energy-momentum tensor is not necessarily conserved. 
However, the conservation of $T_{\mu\nu}$ can be imposed as an extra condition, resulting in the GR equations with an integration constant which has the form of a cosmological term. 
Furthermore, the connection of $\Lambda$ with the the vacuum energy is relaxed, and
conceptually the framework is distinct from the context of GR. It is possible to state that the cosmological constant problem is, at least, alleviated in UG.

Unimodular Gravity admits many generalizations and extensions. For example, the UG constraint can be generalized as $\sqrt{-g} = \xi$, where $\xi$ is interpreted as an external field 
\cite{brand}. This allows one to use any coordinate system, and many new features appear, see for example the discussion of Ref. \cite{espanha} and references therein. 
In Ref. \cite{Kerner} an extension of UG to five dimensions, incorporating the Kaluza-Klein framework, has been implemented. In this version of Kaluza-Klein Unimodular Gravity (KKUG), 
the constraint applies to the five dimensional equations but not necessarily to the reduced equations in four dimensions. The reduction of the five dimensional equations 
to four dimensions leads to a quite unusual structure connecting gravity to a scalar field and ordinary matter, which stems from the five dimensional framework.

Cosmological vacuum solutions in the effective four dimensional equations have been derived in Ref. \cite{Kerner}. They describe a symmetrically bouncing universe. 
There is however a subtle point:  even if the four dimensional metric is free of singularities, the original five dimensional one becomes degenerate at the bounce 
since the modular field associated with the fifth dimension 
vanishes at the transition point from the contracting to the expanding phase. The goal of the present analysis is to show that in spite of the complete regularity in four dimensions, 
the scenario is unstable, reflecting the properties of the five dimensional structure. This analysis will be carried out using tensorial modes in the four dimensional effective theory. 
This limitation does not seem to be a very serious one, since matter is absent from the model, and the effective equations in four dimensions carry the general structure of the original 
equations in five dimensions. 

In order to stress the particular features of the KKUG scenario analyzed here, the same problem is considered in the quantum model in GR on a minisuperspace \cite{lemos} 
where the matter content is the cosmological constant described using the Schutz formalism \cite{schutz1,schutz2}. The resulting scenario is also a symmetric bounce. 
By using an similar classical model \cite{analogue}, it is shown that no signs of instability appear in this case, in opposition to the previous scenario based on the KKUG.

\section{The equations}

Let us briefly remind the formulation of Unimodular Gravity in five dimensions, using the Kaluza-Klein construction, as carried out in Ref. \cite{Kerner}. 
The Kaluza-Klein Unimodular Gravity theory (KKUG) equations read,
\begin{eqnarray}
\label{kkug1}
\tilde R_{AB} - \frac{1}{5}\tilde g_{AB}\tilde R &=& 8\pi G \biggr\{\tilde T_{AB} - \frac{1}{5}\tilde g_{AB}\tilde T\biggl\},\\
\label{kkug2}
\frac{3}{10}{\tilde R}_{;A} &=& 8\pi G\biggr\{{\tilde T}^B_{A;B} - \frac{{\tilde T}_{;A}}{5}\biggl\}.
\end{eqnarray}
The tildes indicate that these equations are written in five dimensions, implying that the indices $A,B$ take values from 0 to 4. 
The field equations (\ref{kkug1}) are traceless. The equation (\ref{kkug2}) expresses the generalized energy-momentum tensor conservation law. 
If the usual conservation law is imposed, meaning ${T^{AB}}_{;B} = 0$, the usual five dimensional equations in presence of the cosmological constant are recovered.

The equations (\ref{kkug1}) can be written in terms of the unimodular gravitational tensor,
\begin{eqnarray}
\tilde E_{AB} = \tilde R_{AB} - \frac{1}{5}\tilde g_{AB}\tilde R,
\end{eqnarray}
and the unimodular matter tensor,
\begin{eqnarray}
\tilde \tau_{AB} = \tilde T_{AB} - \frac{1}{5}\tilde g_{AB}\tilde T, 
\end{eqnarray}
as
\begin{eqnarray}
\tilde E_{AB} = 8\pi G \tilde \tau_{AB}.
\end{eqnarray}

In order to reduce these equations to four dimensions, let us consider the following explicit form of the $5$-dimensional metric tensor:
\begin{eqnarray}
ds_5^2 = g_{\mu\nu}dx^\mu dx^\nu - \phi^2 dx_5^2.
\end{eqnarray}
The components of the five dimensional unimodular gravity tensor become,
\begin{eqnarray}
\label{r1}
\tilde E_{\mu\nu} &=& R_{\mu\nu} - \frac{1}{5}g_{\mu\nu}R - \frac{\phi_{;\mu;\nu}}{\phi} + \frac{2}{5}g_{\mu\nu}\frac{\Box\phi}{\phi},\\
\label{r2}
\tilde E_{55} &=& \frac{ \phi^2}{5}\biggr\{R + 3\frac{\Box\phi}{\phi}.\biggl\}.
\end{eqnarray}
The geometric quantities in the r.h.s. of equations (\ref{r1},\ref{r2}) are constructed using the four-dimensional metric $g_{\mu\nu}$.
We also suppose that the non-vanishing matter terms are only the four dimensional ones. 

The resulting equations are as follows:
\begin{eqnarray}
\label{re1}
R_{\mu\nu} &=& 8\pi G T_{\mu\nu} + \frac{1}{\phi}\biggr(\phi_{;\mu;\nu} - g_{\mu\nu}\Box\phi\biggl),\\
\label{re2}
\frac{\Box \phi}{\phi} &=& \frac{8\pi G}{3}T - \frac{R}{3},\\
\label{re3}
\frac{R_{;\nu}}{2} &=& 8 \pi G\biggr(T^\mu_{\nu;\mu} + \frac{\phi_{;\mu}}{\phi}T^\mu_\nu\biggl).
\end{eqnarray}
Remark that in Ref. \cite{Kerner} there are some misprints in the terms of the matter sector.

The specific properties of the construction displayed here can be verified already in the pure vacuum case. Without matter content, the unimodular field equations read
\begin{eqnarray}\label{TB1}
R_{\mu\nu} &=& \frac{1}{\phi}(\phi_{;\mu;\nu} - g_{\mu\nu}\Box\phi),
\\
\label{TB2}
\frac{\Box\Phi}{\phi} &=& - \frac{R}{3}.
\end{eqnarray}

The vacuum solutions of (\ref{TB1},\ref{TB2}) have been determined in Ref. \cite{Kerner}. They take on the following form:
\begin{eqnarray}
\label{s1}
a(t) &=& a_0\cosh^{1/2}kt,\\
\label{s2}
\phi(t) &=& \phi_0\frac{\sinh kt}{\cosh^{1/2} kt},
\end{eqnarray}
where $k$ is a positive integration constant.
It is a non-singular function with a symmetric bounce around $t = 0$. From now on, we will be using the redefined time coordinate such that $kt \rightarrow t$.
   
This vacuum solution displays the following important property: the strong energy condition is violated during the entire time of evolution of the universe. 
In fact, using (\ref{TB1}) contracted with the four-velocity and inserting the solutions (\ref{s1},\ref{s2}), it comes out that
\begin{eqnarray}
R_{\mu\nu}u^\mu u^\nu = - \frac{3}{2}\frac{1}{\cosh^2 t} \leq 0.
\end{eqnarray}
The strong energy condition is only (marginally) satisfied asymptotically (i.e. for $t \rightarrow \pm \infty$), otherwise it is violated. 
This fact together with the unusual form of the four-dimensional field equations allow us to expect that the solutions (\ref{s1},\ref{s2}) display some singularities. In what follows we will show explicitly that this is the case showing 
that the gravitational perturbations do indeed diverge around the bounce.

\section{Perturbative analysis}

The perturbation of the field equations (\ref{TB1}) reads:
\begin{eqnarray}
\delta R_{\mu\nu} = - \frac{1}{\phi^2}(\phi_{;\mu;\nu} - g_{\mu\nu}\Box\phi)\delta\phi + \frac{1}{\phi}\biggr(\delta \phi_{;\mu\nu} 
- h_{\mu\nu}\Box\phi - g_{\mu\nu} \delta \Box\phi\biggl).
\end{eqnarray}
From now on, the synchronous coordinate condition, $h_{\mu\nu} = 0$, will be used. In this {\it gauge}, the non-zero perturbations 
of the Christoffel symbols, $\chi^\rho_{\mu\nu}$, become \cite{wei},
\begin{eqnarray}
\chi^0_{ij} &=& - \frac{\dot h_{ij}}{2},\\
\chi^i_{0j} &=& - \frac{1}{2}\biggr(\frac{h_{ij}}{a^2}\biggl)^\bullet,\\
\chi^i_{jk} &=& - \frac{1}{2a^2}\biggr\{\partial_j h_{ik} + \partial_k h_{ij} - \partial_i h_{jk}\biggl\}.
\end{eqnarray}
The perturbed Ricci tensor reads,
\begin{eqnarray}
\delta R_{\mu\nu} = \partial_\rho\chi^\rho_{\mu\nu} - \partial_\nu\chi^\rho_{\mu\rho} + \Gamma^\rho_{\sigma \rho}\chi^\sigma_{\mu\nu} 
+ \Gamma_{\mu\nu}^\sigma\chi^\rho_{\sigma\rho} - \Gamma_{\mu\rho}^\sigma\chi^\rho_{\sigma\nu}
- \Gamma^\sigma_{\nu\rho}\chi^\rho_{\sigma\mu}.
\end{eqnarray}

With the definition
\begin{eqnarray}
h = \frac{h_{kk}}{a^2},
\end{eqnarray}
the components of the perturbed Ricci tensor read \cite{wei}, 
\begin{eqnarray}
\delta R_{00} 
&=& \frac{\ddot h}{2} + H\dot h,\\
\delta R_{0i} &=& \frac{1}{2}\biggr\{\partial_i \dot h - \frac{\partial_k \dot h_{ki}}{a^2} + 2 H\frac{ \partial_k h_{ki}}{a^2}\biggl\},\\
\delta R_{ij} &=& - \frac{\ddot h_{ij}}{2} + \frac{H}{2}\biggr(\dot h_{ij} - \delta_{ij}\dot h_ {kk}\biggl) - 2H^2h_{ij} + \delta_{ij}H^2h_{kk}\nonumber\\
&-& \frac{1}{2a^2}\biggr\{\partial_j\partial_k h_{ik} + \partial_i\partial_k h_{kj} - \nabla^2h_{ij} - \partial_i\partial_j h_{kk }\biggl\}.
\end{eqnarray}
The perturbed Ricci scalar is given by
\begin{eqnarray}
\delta R &=& \ddot h + 4H\dot h - \frac{\nabla^2h}{a^2} + \frac{\partial_k\partial_l h_{kl}}{a^4}.
\end{eqnarray}

The perturbations of the energy-momentum tensor
\begin{eqnarray}
T^{\mu\nu} = (\rho + p)u^\mu u^\nu - p g^{\mu\nu}.
\end{eqnarray}
are given by,
\begin{eqnarray}
\delta T^{00} &=& \delta\rho,\\
\delta T^{i0} &=& (\rho + p)\delta u^i,\\
\delta T^{ij} &=& h^{ij}p - g^{ij}\delta p.
\end{eqnarray}

\section{Gravitational waves}

The tensorial modes, related to gravitational waves, are obtained by retaining the transverse, traceless components of $h_{ij}$ and fixing $\delta\phi = 0$, 
since it contributes only to the scalar modes. The resulting equation is,
\begin{eqnarray}
\ddot h_{ij} - \biggr(H - \frac{\dot\phi}{\phi}\biggl)\dot h_{ij} + \biggr\{\frac{q^2}{a^2} + 4H^2 
- 2\biggr(\frac{\ddot\phi}{\phi} + 3H\frac{\dot\phi}{\phi}\biggl)\biggl\}h_{ij} = 0.
\end{eqnarray}
After inserting the background solutions, the equation for gravitational waves reads,
\begin{eqnarray}
\label{gwe}
\ddot h_{ij} + \frac{\dot h_{ij}}{\cosh t\sinh t} + \biggr\{\frac{q^2}{\cosh t} - \frac{1}{\cosh^2 t} - 1\biggl\}h_{ij} = 0.
\end{eqnarray}
This equation does not seem to admit an exact solution. Hence, we must proceed by performing some asymptotic expansion complementing the analysis with 
numerical computation.

The bounce occurs at $t  = 0$. Near the bounce, the equation takes the form,
\begin{eqnarray}
\ddot h_{ij} + \frac{\dot h_{ij}}{ t} + \biggr\{q^2 - 2\biggl\}h_{ij} = 0.
\end{eqnarray}
Defining $\tilde q^2 = q^2 - 2$, the solutions read,
\begin{eqnarray}
h_{ij} = \epsilon_{ij}\biggr\{c_1 J_0 (\tilde q t) + c_ 2 N_0(\tilde q t)\biggl\}.
\end{eqnarray}
In this expression $c_{1,2}$ are integration constants and $J_0$ and $N_0$ are Bessel's and Neumann's functions of order zero. 
Remark that the Neumann function diverges logarithmically for
$t = 0$. If $\tilde q^2 < 0$, the solutions are written in terms of modified Bessel functions $K_0$ and $I_0$. There is again a singularity 
associated with the function $K_0$.

In the asymptotic regions, given by $t \rightarrow \pm \infty$, the equation reduces to, 
\begin{eqnarray}
\ddot h_{ij} - h_{ij} = 0,
\end{eqnarray}
with the solution,
\begin{eqnarray}
h_{ij} = c_-e^{-t} + c_+t^{t}.
\end{eqnarray}
In order to have finite initial conditions, such that the perturbative study is justified, $c_- = 0$ if the initial conditions are imposed at $t \rightarrow - \infty$.

There is one important point. From equation (\ref{gwe}), the condition for the propagation of the gravitational waves is given by,
\begin{eqnarray}
\omega^2 = \frac{q^2}{\cosh t} - \frac{1}{\cosh^2 t} - 1 > 0.
\end{eqnarray}
Hence, the propagation of gravitational waves is possible only for a positive time $t_0$ on, given by the following expression:
\begin{eqnarray}
t_0 = \frac{q^2}{2} \pm \sqrt{\biggr(\frac{q^2}{2}\biggl)^2 - 1}.
\end{eqnarray}
Since $t_0 > 0$, no mode satisfying this condition exist during the contracting phase. Modes whose wave numbers satisfy $q^2 < 2$ do not propagate during all times between $t= -\infty$ and $t=+\infty$.
In the expanding phase, only modes with $q^2 \geq 2$ can propagate. 
Even if a given mode does not propagate, the perturbations still grow with time. Hence, the configuration is unstable under gravitational perturbations. 

\begin{figure}[!ht]
	\centering
	\includegraphics[width=0.45\columnwidth]{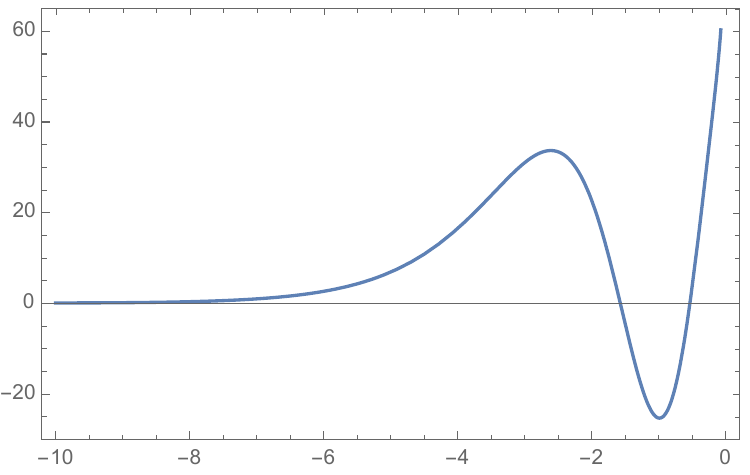}
	\includegraphics[width=0.45\columnwidth]{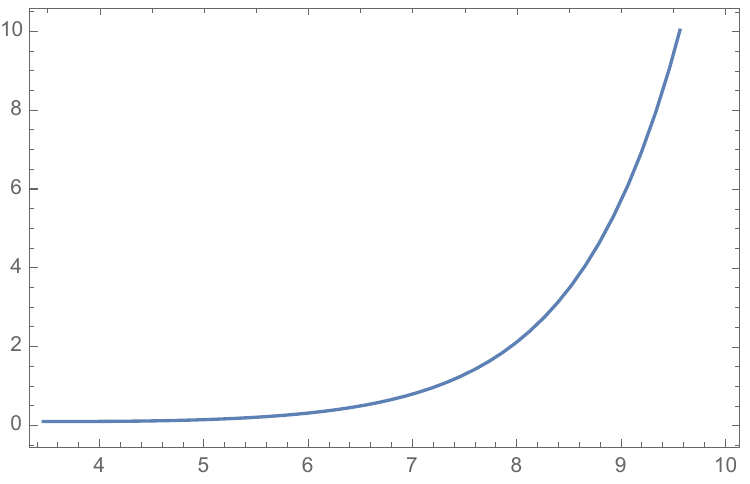}
	\caption{Behavior of the perturbations for the contracting phase (left panel) and expanding phase (right panel). In both cases, $q = 4$.}
	\label{fig1}
\end{figure}

\begin{figure}[!ht]
	\centering
	\includegraphics[width=0.45\columnwidth]{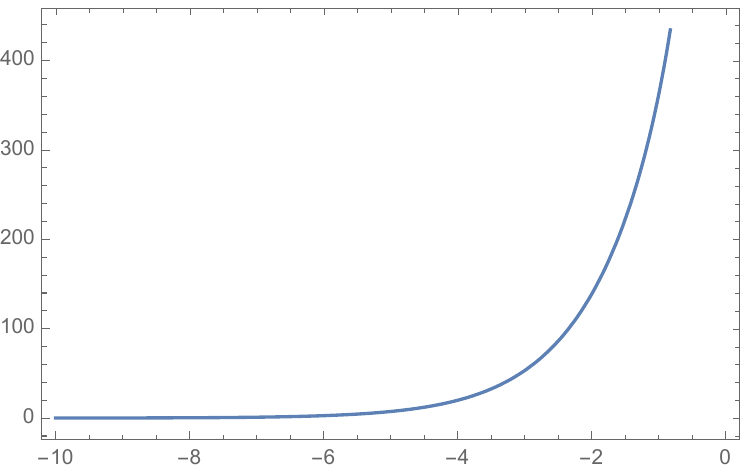}
	\includegraphics[width=0.45\columnwidth]{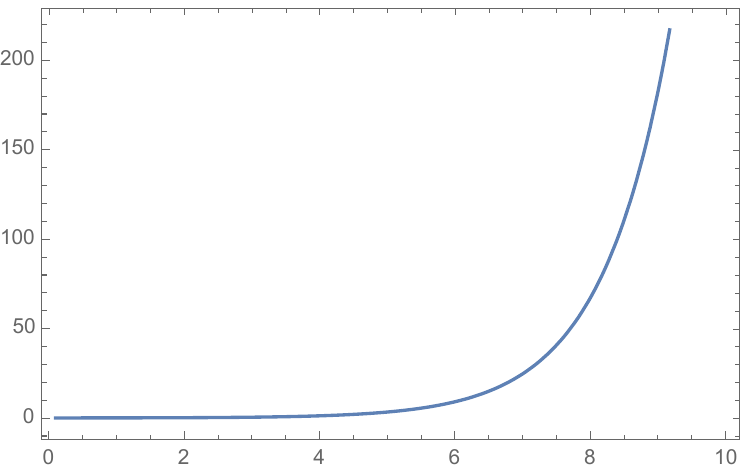}
	\caption{Behavior of the perturbations for the contracting phase (left panel) and expanding phase (right panel). In both cases, $q = 1$.}
	\label{fig2}
\end{figure}

As a matter of fact, the modes in the contracting and in the expanding phases
are disjoint due to the divergence at the bounce at $t = 0$. In figures \ref{fig1} and \ref{fig2}, the divergences of perturbations in the contracting phase 
and their behavior during the expanding phase of the evolution of the universe are displayed for two different values of the wavenumber $q$. 
 
\section{A non singular solution with cosmological constant}

In Ref. \cite{lemos} a quantum cosmological model using General Relativity theory and employing the Schutz formalism to describe the matter content
has been analyzed. General solutions were found for the linear equation of state $p = \alpha\rho$. The quantum effects lead to a singularity free solution. 
Here we will concentrate on the cosmological constant case given by $\alpha = - 1$. The reason for this choice is that in this case the system admits, firstly, 
an analytical solution in terms of the cosmic time $t$ similar to the one found above. Secondly, UG has in general a hidden connection 
with a cosmological constant, what motivates us to explore this particular case. In the Ref. \cite{lemos} a general solution for the expectation value
of the scale factor was found. However, for the present purpose we will use a classical analogue model \cite{analogue}. The reason is that the perturbative 
analysis, even for gravitational waves, becomes quite entangled with a full quantum model, and it is more convenient to work with this classical analogue 
such that the background solution is equivalent to the quantum expectation value.

The classical analogue model discussed in Ref. \cite{analogue} establishes that if a stiff matter fluid with negative energy is added to a given fluid, 
the scale factor exhibits the same behavior as the expectation value for the quantum model in the mini-superspace with the fluid described 
by the Schutz formalism. Hence, for a cosmological constant, the analogue classical model is given by the equation
\begin{eqnarray}
H^2 = \frac{8\pi G}{3}\biggr(\rho_{\Lambda0} - \rho_{s0}a^{-6}\biggl),
\end{eqnarray}
where the $\rho_{\Lambda0}$ and $\rho_{s0}$ are the cosmological constant and stiff matter density today. The energy conditions are also violated in all the evolution
of the universe for this model. The solution reads,
\begin{eqnarray}
\label{scc}
a(t) = \biggr(\frac{\Omega_{s0}}{\Omega_{\Lambda0}}\biggl)^{1/6}\cosh ^{1/3}(3\sqrt{\Omega_{\Lambda0}}t).
\end{eqnarray}
The fractional (partial) densities are given by,
\begin{eqnarray}
\Omega_{\Lambda0} = \frac{8\pi G}{3H_0^2}\rho_{\Lambda0}, \quad \Omega_{s0} = \frac{8\pi G}{3H_0^2}\rho_{s0}.
\end{eqnarray}

The solution (\ref{scc}) represents a symmetric bounce similar to that found in KKUG. Using the expressions for the perturbations of the previous section, 
it is quite direct to compute the equation governing the evolution of gravitational waves for the cosmological constant model. It reads,
\begin{eqnarray}
\ddot h_{ij} - H\dot h_{ij} + \biggr\{\frac{q^2}{a^2} - 2(\dot H + H^2)\biggl\}h_{ij} = 0.
\end{eqnarray}
It is convenient to perform the following rescaling:
\begin{eqnarray}
3\sqrt{\Omega_{\Lambda0}}t \rightarrow t, \quad \frac{q}{3\sqrt{\Omega_{\Lambda0}}} \rightarrow q.
\end{eqnarray}
The Hubble function $H$ reads,
\begin{eqnarray}
H = \frac{1}{3}\tanh t.
\end{eqnarray}

\begin{figure}[!ht]
	\centering
	\includegraphics[width=0.45\columnwidth]{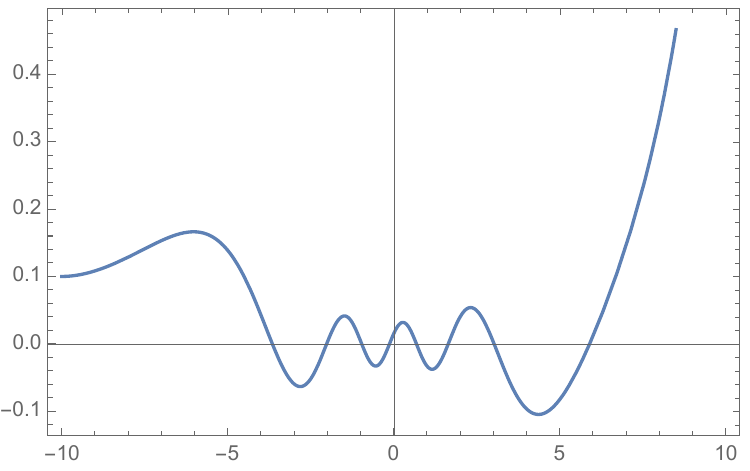}
	\includegraphics[width=0.45\columnwidth]{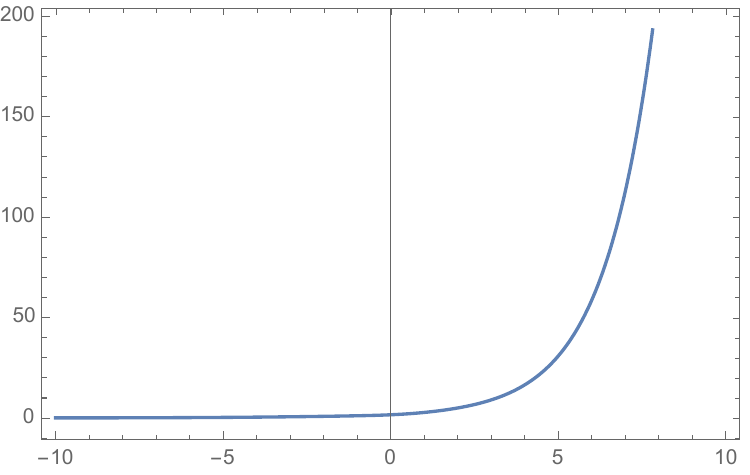}
	\caption{Behaviour of the perturbations for $q = 4$ (left panel) and $q = 0.5$ (right pannel).}
	\label{fig3}
\end{figure}

In figure \ref{fig3} the behavior of gravitational waves is displayed for two values of $q$. In opposition to what was found previously, there is a regular behavior 
with no explicit sign of instability.
Only asymptotically, for large $t$, a divergence persists. However, this means that the perturbations are growing, and from a certain moment the linear approximation used here 
is not valid anymore. In particular, in opposition to the KKUG, the perturbations behave regularly in the bounce, and there no disjoint regions (corresponding to the 
contracting and expanding phases) appear as what is observed in the KKUG case.

The numerical result can be compared with the exact solution in the asymptotic regions.
For $t \rightarrow 0$ (the bounce), the equation reduces to, 
\begin{eqnarray}
\ddot h_{ij}  - \frac{t}{3}\dot h_{ij} + \biggr\{q^2 - \frac{2}{3}\biggl\}h_{ij} = 0.
\end{eqnarray}
This equation can be solved in terms of confluent hypergeometric functions which reduces, very near the bounce to a constant mode and a linearly increasing function of $t$: The solutions are regular at the bounce, being continuous while traversing from the contracting to the expanding phases..

 For $t \rightarrow \pm \infty$, the equation reads,
\begin{eqnarray}
\ddot h_{ij} \mp \frac{\dot h_{ij}}{3} - \frac{2}{9}h_{ij} = 0.
\end{eqnarray}
The solutions are for $t \rightarrow \pm \infty$,
\begin{eqnarray}
h_{ij} \propto \epsilon_{ij}t^{p_{1,2}},
\end{eqnarray}
with $p_1 = \pm 2/3$ and $p_2 = \mp 1/3$, the superior sign corresponding to $t \rightarrow \infty$ and the inferior one to $t \rightarrow - \infty$.
Asymptotically, two modes do appear, a growing and a decaying one. Due to the time reversal symmetry of the model, the growing and decaying modes 
change their rôle when one passes from one asymptotic to another, as it could be expected from the bounce symmetric with respect to time reversal.

\section{Concluding remarks}

The reduction of the Unimodular Kaluza-Klein to four dimensions leads to a very peculiar gravitational system which is similar to the Brans-Dicke theory with vanishing potential 
and $\omega = 0$ \cite{Brans}, but with the crucial difference which is that
the Ricci scalar is absent from the geometric sector, see equations (\ref{re1},\ref{re2},\ref{re3}). The vacuum solutions display a symmetric bounce, completely regular 
from the four dimensional point of view. Remark that symmetric and asymmetric bounces in general have very different properties, see Ref. \cite{nelson} for discussion. 
Here, we have shown that these solutions are unstable. The analysis were made using the tensorial modes in a linear perturbative approach. This result has been compared 
with a solution displaying very similar features (in particular, also a symmetric bounce) but coming from a quantum model in four dimensions, which reveals to be stable. 

What is the reason for this instability? We must remember that the original five-dimensional theory is singular, since the field $\phi$ is zero at the bounce: 
$\phi$ is connected with the fifth dimension, and at the bounce the five-dimensional metric becomes degenerate. The structure of equations (\ref{re1},\ref{re2},\ref{re3}) 
seems to reflect in someway this situation. Of course, in view of this, it is possible to ask a perturbative analysis in the original five-dimensional structure 
is much more cumbersome due to the fact that the five-dimensional metric is anistropic. But, since vacuum solutions are considered modes that appear in the four-dimensional model 
will also appear in five dimensions and their instabilities is enough to state that the five-dimensional configuration is unstable.

It should be stressed that we are considering exclusively vacuum solutions and the presence of matter may modify the above results.  
Ordinary matter may change the energy conditions, which for vacuum are always violated. It is not possible to attribute the instability to the violation of the energy conditions 
during the entire evolution of the universe since this violation occurs also for the quantum effective model considered here, and which reveals to be stable. 
It may nevertheless bring new features to the solutions avoiding the degeneracy of the five dimensional geometry. We hope to be able to present this analysis in a separate work.

\bigskip

\noindent
{\bf Acknowledgments:} JCF thanks CNPq (Brasil) and FAPES (Brazil) for financial support.

\end{document}